\documentclass{aa}
\usepackage{txfonts}
\usepackage{graphicx}

\newcommand{\ub}{\mbox{$U\!-\!B$}}
\newcommand{\bv}{\mbox{$B\!-\!V$}}

\begin{document}
\title{Planetary Nebula Candidates in Extragalactic Young Star Clusters
  \thanks{Based on observations collected at the European Southern
          Observatory, Chile under programme 072.B-0242(A)
	  and with the NASA/ESA
    \emph{Hubble Space Telescope}, obtained at the Space Telescope
	  Science Institute, which is operated by the Association
	  of Universities for Research in Astronomy, Inc., under
	  NASA contract NAS 5-26555}
}

\author{S. S. Larsen \inst{1,2} \and T. Richtler \inst{3}}

\institute{European Southern Observatory, ST-ECF,
  Karl-Schwarzschild-Str. 2, D-85748 Garching b.\ M{\"u}nchen, Germany
  \and
  Astronomical Institute, University of Utrecht, Princetonplein 5,
  NL-3584 CC, Utrecht, The Netherlands
  \and
  Astronomy Group,
  Universidad de Concepci{\'o}n, Departamento de F{\'i}sica,
  Casilla 160-C, Concepci{\'o}n, Chile
}

\offprints{S.\ S.\ Larsen, \email{larsen@astro.uu.nl}}

\date{Received 13 June 2006 / Accepted 16 August 2006}

\abstract
{
During an analysis of optical spectra of 80 young star clusters in several 
nearby spiral galaxies, $[$\ion{O}{iii}$]$ and $[$\ion{N}{ii}$]$ emission 
lines were noted in some cases.  Three of these emission
line sources are associated with clusters older than 30 Myrs, and are 
identified as likely planetary nebula (PN) candidates.
}
{
These objects may represent a rare opportunity to study PNe whose progenitor 
stars are known to be of intermediate masses, although detailed analysis is 
challenging because of the 
underlying strong continuum from the cluster stars. This paper presents
and discusses basic properties of the PN candidates and their host clusters.
}
{
Based on the observed emission line fluxes, the excitation parameters and
luminosities of the nebulae are derived. This allows a crude placement of
the central stars in two of the objects on the H-R diagram. Cluster ages
and masses are estimated from broad-band colours and by fitting model
SSP spectra to the observed spectra.
}
{
The two PN candidates where central star luminosities and temperatures 
can be estimated 
are found to be consistent with post-AGB model tracks for a central star 
mass of about 0.60 $M_{\odot}$.  
One of the host clusters has an age of 32--65 Myrs, 
corresponding to a main sequence turn-off mass of 
$M_{\rm TO}$ = 6.6--9.0 M$_\odot$. For the other cluster the age is 
282--407 Myrs, corresponding to M$_{\rm TO}$ = 3.2--3.6 M$_\odot$.
By estimating the number of stars evolving off the main
sequence per year, a total of 6 PNe are expected in the full sample of
80 clusters
for a PN lifetime of $10^4$ years. The factor of two disagreement with 
the actual observed number may be due, among other things, to uncertainties
in PN lifetimes. It is interesting to note that all three PN
candidates are associated with clusters which are more diffuse than
average.
}
{
While PNe have previously been found in some old globular clusters, the 
candidates identified here are among the first identified 
in \emph{young} star clusters. 
}


\titlerunning{Planetary Nebula Candidates in YMCs}
\maketitle

\section{Introduction}

Star clusters and planetary nebulae (PNe) can both be observed at distances
well beyond the Local Group, and are thus valuable extragalactic population 
tracers. However, much remains to be understood about the post-main sequence 
evolution of low- and intermediate stars leading up to the PN stage, partly 
because of the difficulty in determining the distances and progenitor
stellar masses of planetary nebulae in the Milky Way. For PNe 
associated with stellar clusters the initial mass of the progenitor star can 
be assumed to be that corresponding to the main sequence turn-off (unless 
the progenitor star is a close binary in which mass exchange has occurred),
but the number of PNe known to be associated with stellar clusters 
remains small. Less than a handful of such objects are known to exist in
Milky Way open clusters (Pedreros \cite{ped87,ped89}; O'Dell \cite{odell63})
and four PNe have been found in Galactic globular clusters 
(M15, M22, Pal 6, NGC~6441; Jacoby et al.\
\cite{jac97}). 

The distance problem can be alleviated by studying
PNe in external galaxies with well determined distances. Many such
searches have been carried out or are currently under way 
(Kwok \cite{kwok00}; Feldmeier \cite{feld06}; 
Magrini \cite{mag06}), but the detected
candidates are typically field objects and the initial mass thus
remains unknown. PNe have been identified in a few GCs in the nearby
giant elliptical NGC~5128 (Minniti \& Rejkuba \cite{mr02}; Rejkuba et al.\
\cite{rej03}), and the significant number of on-going large
spectroscopic surveys of extragalactic globular cluster systems are
likely to reveal many more candidates. However, by studying PNe in 
ancient globular clusters one remains fundamentally limited to 
low-mass stars. One way to sample a larger range of
stellar masses would be to collect a sample of PN candidates in younger
stellar clusters. So far, the number of such objects are limited to the
small number of PNe in Milky Way open clusters.

We have recently obtained optical spectra for a number of young star
clusters in several nearby galaxies. The original aim of this programme was
to carry out a detailed test of simple stellar population (SSP) models,
but during the analysis we noted
emission lines in some of the spectra. As discussed below, the 
clusters in question span a range of ages from a few tens of Myrs to several 
hundreds of Myrs, corresponding to main sequence turn-off masses between 3 
and 9 M$_{\odot}$.  These objects potentially represent a valuable addition 
to the 
very few cases where PNe have been associated with young star clusters, and 
could provide a unique opportunity to study PNe where the 
distances and initial masses 
are known with some accuracy.  The full spectroscopic dataset and analysis
will be presented in a forthcoming paper, and here we concentrate on the 
few spectra where PN candidates were identified.

\section{Data}

Spectra of a sample of young massive clusters (YMCs) in several nearby spiral 
galaxies were obtained with the EMMI instrument on the 
ESO New Technology Telescope from March 16 - March 18, 2004. The observations 
were carried out in multi-slit mode, using the RILD (``Red Imaging and 
Low-Dispersion Spectroscopy'') mode and grism \#3, which provided a spectral 
resolution of $\lambda/\Delta\lambda\sim760$ over the wavelength range 
4000\AA -- 9000\AA .  In each of the 7 slitmasks, typically 3 exposures of 
3600 s each were obtained with the CCDs read out in 2$\times$2 binned mode.  
The full spectroscopic sample included 80 clusters in four galaxies 
(NGC~2835, NGC~2997, NGC~3621 and NGC~5236), selected primarily from the sample 
presented in Larsen (\cite{lar99}).  In order to maximise the fraction of
bona-fide star clusters (as opposed to ``asterisms'' or loose
associations), as many objects as possible with imaging from the Hubble Space 
Telescope (HST) were included (Larsen \cite{larsen04}). 
The astrometry for the slitmask design was based on pre-imaging data obtained 
with EMMI in service mode on Jan 28, 2004, exposed for 5 min in the $R$-band.

The spectra were reduced with standard tools in the 
IRAF\footnote{IRAF is distributed by the National Optical Astronomical
Observatories, which are operated by the Association of Universities for
Research in Astronomy, Inc.~under contract with the National Science
Foundation}
ONEDSPEC package
and were wavelength- and flux calibrated using observations of HeAr
calibration lamps and the flux standards Hiltner 600 and LTT 2415
(Stone \& Baldwin \cite{sb83}; Hamuy et al.\ \cite{ham94}). Small additional 
corrections (on order $\sim1$ \AA) to the wavelength 
scale were applied using the [\ion{O}{i}] 5577.35\AA\ night sky line.

Accurate absolute flux calibration of slit spectra is always challenging due 
to slitlosses,
which depend on the seeing and accurate centring of objects on
the slit. Because EMMI does not have a functioning atmospheric dispersion
corrector, an additional complication for multislit spectroscopy is
differential refraction, which causes the slitlosses to be wavelength
dependent as the slits cannot be aligned with the parallactic
angle for the entire duration of an exposure. Other factors, such as the
spatial distribution of the targets, also limit the freedom to choose the
orientation of the slits. Although the observations were carried out at
low airmass (generally $<1.2$), some differences were noted
in the flux calibration between individual observations of the same
targets.

We attempted to bring our spectra closer to an absolute flux scale by scaling 
them to match the fluxes derived from the ground-based photometry. Since the 
wavelength range covered by the EMMI spectra includes four of the photometric
passbands ($BVRI$) for which broad-band photometry is available, the 
wavelength-dependent
nature of the slit losses could also be accommodated. This was done
by forming the ratio of the photometric fluxes ($F_{\rm phot}(\lambda)$) 
in each band to 
the fluxes measured in the corresponding wavelength ranges of the 
spectra ($F_{\rm spectro}(\lambda)$), and then producing
a correction function by fitting a second-order polynomial to the
resulting ratio $F_{\rm phot}/F_{\rm spectro}(\lambda)$.  Within a 
given exposure, differences in the overall scaling of individual spectra
could be as large as $\sim25$\%.  This may reflect uncertainties in the 
photometric as well as the spectroscopic measurements.  The \emph{shapes} of 
the individual correction functions showed better agreement.  Furthermore, 
for some objects (e.g.\ \ion{H}{ii} regions dominated by strong emission 
lines) it was not possible to construct reliable 
correction functions. We therefore applied
one average correction function to all spectra obtained in
a given exposure, based on about half of the objects in a given
slitmask.
Typically, the correction amounted to 10\%--20\%, but
in the most extreme cases it could be as much as a factor of 2. 
Each spectrum was then multiplied by the correction 
function, and the several individual spectra of each object were
averaged (using a sigma-clipping algorithm to reject cosmic-ray hits)
to produce the final spectra.

Visual inspection of the reduced spectra generally revealed A-type
spectra with strong Balmer lines, as expected for young star clusters
with ages on the order of $\sim 10^8$ years.  Some very young objects, known 
from ground-based narrow-band imaging to be embedded in H$\alpha$ emission,
were also included, and showed the expected strong emission 
lines from H, N, and O characteristic of \ion{H}{ii} regions.  Here we 
concentrate on three 
spectra which showed weaker [\ion{O}{iii}] and [\ion{N}{ii}] emission lines 
superimposed on A-type stellar spectra, for which both broad-band photometry
and a comparison with model 
spectra suggest ages significantly greater than $10^7$ 
years (\S\ref{sec:hostprops}).  Ground-based H$\alpha$ images showed no 
line emission in the
immediate vicinity of these objects, again suggesting that the emission-line
objects contained in the star clusters are not \ion{H}{ii} regions. As 
discussed below, a more likely (although not entirely 
unique) explanation is that these clusters contain 
planetary nebulae (PNe). 

The three clusters discussed here are located in the two galaxies 
NGC~5236 and NGC~3621, both of which are actively star-forming, nearby spiral 
galaxies.  Throughout this paper, we assume distance moduli of 
$m-M=28.25\pm0.15$ for NGC~5236 (Thim et al.\ \cite{thim03}) and 
$29.10\pm0.20$ for NGC~3621 (Rawson et al.\ \cite{raw97}). For the foreground 
extinction we use the Schlegel et al.\ (\cite{sch98}) values of 
$A_B=0.284$ mag (NGC~5236) and $A_B=0.346$ mag (NGC~3621).
For a more comprehensive listing of the properties of the host galaxies we
refer to Larsen \& Richtler (\cite{lr00}).

\section{Analysis}

\subsection{\bf Basic properties of the emission-line sources}

\begin{figure}
\includegraphics[width=85mm]{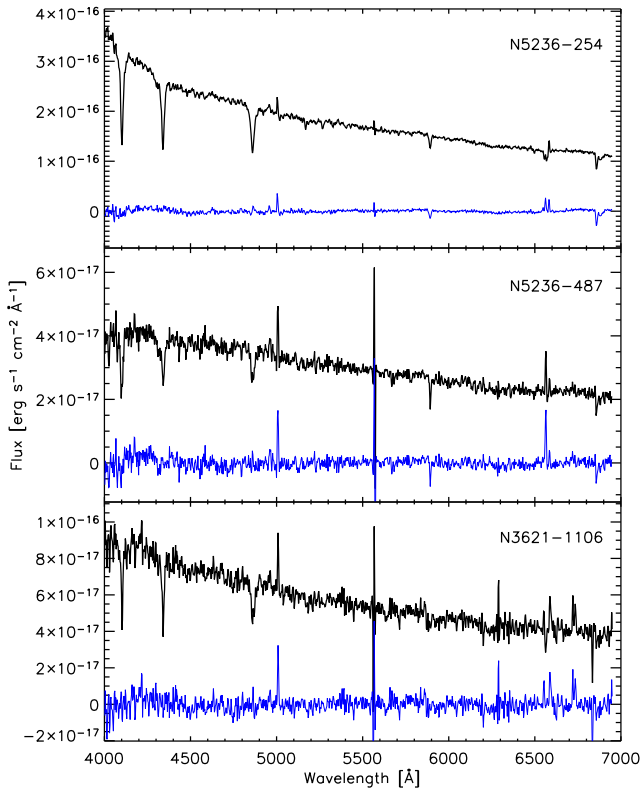}
\caption{EMMI spectra of the three YMCs with PN candidates. Each panel shows
 the raw spectrum (thick lines) and the residuals when the 
 best-fitting solar-metallicity
 model spectrum from Gonz{\'a}lez-Delgado et al.\ (\cite{gon05}) is 
 subtracted (blue lines).}
\label{fig:spec}
\end{figure}

\begin{figure}
\begin{minipage}{28mm}
N5236-254 \\
\includegraphics[width=28mm]{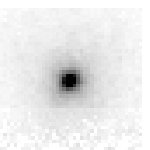}
\end{minipage}
\begin{minipage}{28mm}
N5236-487 \\
\includegraphics[width=28mm]{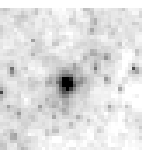}
\end{minipage}
\begin{minipage}{28mm}
N3621-1106 \\
\includegraphics[width=28mm]{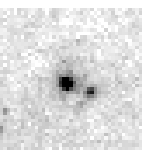}
\end{minipage}
\caption{HST/WFPC2 F606W ($\approx V$-band) images of the three YMCs. 
Each panel is $4\arcsec\times4\arcsec$ and the scale is $0\farcs10$
pixel$^{-1}$.}
\label{fig:stamps}
\end{figure}

\begin{table}
\caption{\label{tab:emlines}Emission line fluxes in units of 
  $10^{-17}$ erg s$^{-1}$ cm$^{-2}$. Note that both
  lines in the $[$\ion{O}{iii}$]$ and $[$\ion{N}{ii}$]$ doublets
  were fitted together.}
\begin{tabular}{lccc} \hline
			     & \multicolumn{3}{c}{Fluxes $[10^{-17}$ erg s$^{-1}$ cm$^{-2}$ $]$} \\
                             &  N5236-254   & N5236-487    & N3621-1106 \\
Line                         &              &              &            \\
H$\beta$ 4861                &   6.0        &  2.2         &  4.6       \\
H$\alpha$ 6562               &  18.4        & 12.3         & -0.15      \\
$[$\ion{O}{iii}$]$ 4959,5007 &   6.9, 20.6  &  4.1, 12.1   &  7.2, 21.7 \\ 
$[$\ion{N}{ii}$]$ 6548,6584  &   5.6, 16.7  &  0.7,  2.2   &  4.7, 14.1 \\
Mean Error                   &   1.3        &  0.8         &  2.9       \\ 
$m_{5007}$              & $25.5\pm0.06$ & $26.1\pm0.07$ & $25.4\pm0.13$ \\
\hline
\end{tabular}
\end{table}

The EMMI spectra of the three objects are shown in Fig.~\ref{fig:spec}. 
All three objects happened to have images from the Wide Field Planetary 
Camera 2 (WFPC2) on board the Hubble Space Telescope (HST), as shown in 
Fig.~\ref{fig:stamps}. Two of
the objects are located in NGC~5236 (N5236-254, N5236-487 in the
list of Larsen \cite{lar99}), while the third is in NGC~3621
(N3621-1106). The morphological appearance of all three objects 
is generally consistent with that of star clusters, all of them being 
significantly more extended than the WFPC2 point-spread function, but still 
fairly compact and with regular, symmetric profiles. Note, though, that
N3621-1106 has a fainter companion at a projected separation of
$0\farcs7$ ($\sim22$ pc). The companion is also resolved in the HST 
images, and thus also a likely cluster candidate. Whether the two objects 
represent a chance alignment or a real physical pair is difficult
to tell, however. Similar colours would suggest similar ages for the
two objects and thus make a chance alignment less likely, but the ground-based 
imaging has insufficient resolution to address this question and the HST 
data are available in only one band. The half-light radii listed 
in Larsen (\cite{larsen04}) are 10.1 pc, 6.7 pc and 7.2 pc for N5236-254,
N5236-487 and N3621-1106, respectively. While these sizes are
consistent with those of star clusters, it is interesting to note that
all three clusters have larger half-light radii than the mean
value of 3--4 pc which is typical for both young and old clusters
(Whitmore et al.\ \cite{whit99}; Larsen \cite{larsen04}; 
 Jord{\'a}n et al.\ \cite{jordan05}).

  The spectral signal-to-noise ratio per pixel in the dispersion direction is 
about 106, 30 and 24 at 5000\AA\ for N5236-254, N5236-487 and N3621-1106, 
respectively.  The [\ion{O}{iii}] $\lambda$4959,5007\AA\ and 
[\ion{N}{ii}] $\lambda$6548,6584\AA\ lines are noticeable in all three
spectra, but much weaker than typically seen in \ion{H}{ii}
regions. Balmer lines are expected to be seen in emission in both
planetary nebulae and \ion{H}{ii} regions, but in our cluster spectra
the Balmer lines are dominated by the underlying strong absorption 
lines from the stellar component. 

In order to remove the stellar component from the spectra, model spectra from 
the library of Gonz{\'a}lez-Delgado et al.\ (\cite{gon05}) were subtracted 
from the observed spectra.  
The library contains model spectra for the
integrated light of simple stellar populations with ages between
1 Myr and 17 Gyr, tabulated at 0.3\AA\ resolution. A Salpeter 
(\cite{salp55})-like IMF is assumed.
The best-fitting model spectra were selected by 
minimising the r.m.s.\ residuals between the observed spectra (shifted
to zero radial velocity) and smoothed model spectra.  Because of the
uncertainties in the absolute flux calibration and the reddening
corrections, the model spectra were multiplied by a polynomial to
follow the same large-scale spectral energy distribution as the
observed spectra.
This was done by fitting the ratio of the observed to model spectra
with a 5th-order polynomial and multiplying the model spectra with
this polynomial, rather than scaling them by a single constant, before 
calculating the r.m.s.  residuals.  The Gonz{\'a}lez-Delgado et al.\
library includes
models based on both the Geneva and Padua stellar isochrones and for a range 
of metallicities, but for this work we only used Solar metallicity models 
based on the Padua isochrones.  The analysis carried out in the following
does not depend strongly on the model choice. 
For NGC~5236, analysis of \ion{H}{ii} regions 
(Bresolin \& Kennicutt \cite{bk02}) yields an oxygen
abundance very close to Solar [12+log(O/H)=8.9] at the galactocentric 
distance of the two clusters ($6\farcm3$ for N5236-254 and $5\farcm1$ for 
N5236-487). For NGC~3621, the work of Ryder (\cite{ryder95}) indicates
12+log(O/H)=9.0 at the position of N3621-1106, $1\farcm0$ from the centre.
The use of Solar metallicity models thus appears justified. The only 
remaining free parameter is then the age, and for each cluster we selected 
the model spectrum of the age which gave the best fit to the data.

\begin{figure}
\includegraphics[width=85mm]{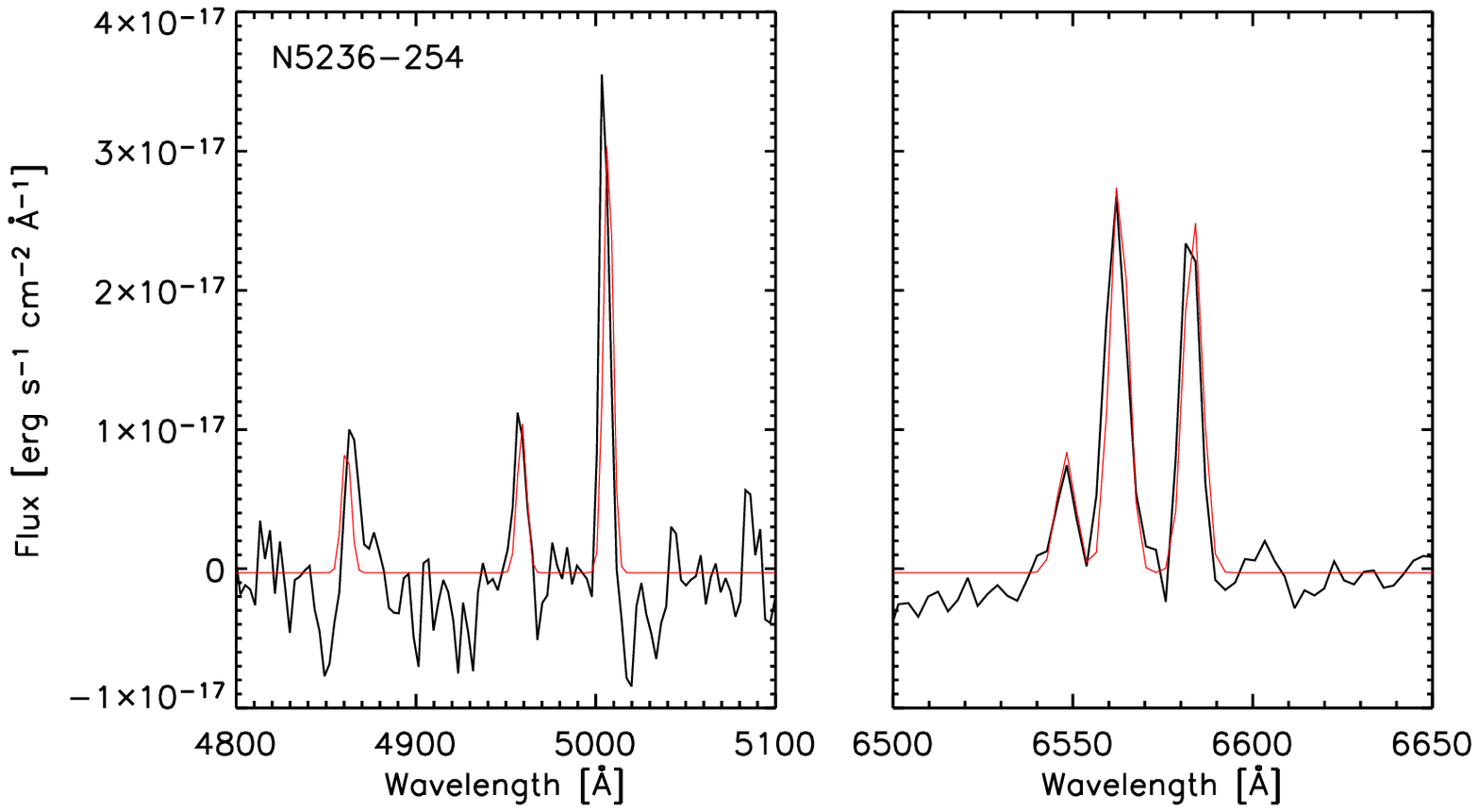}
\includegraphics[width=85mm]{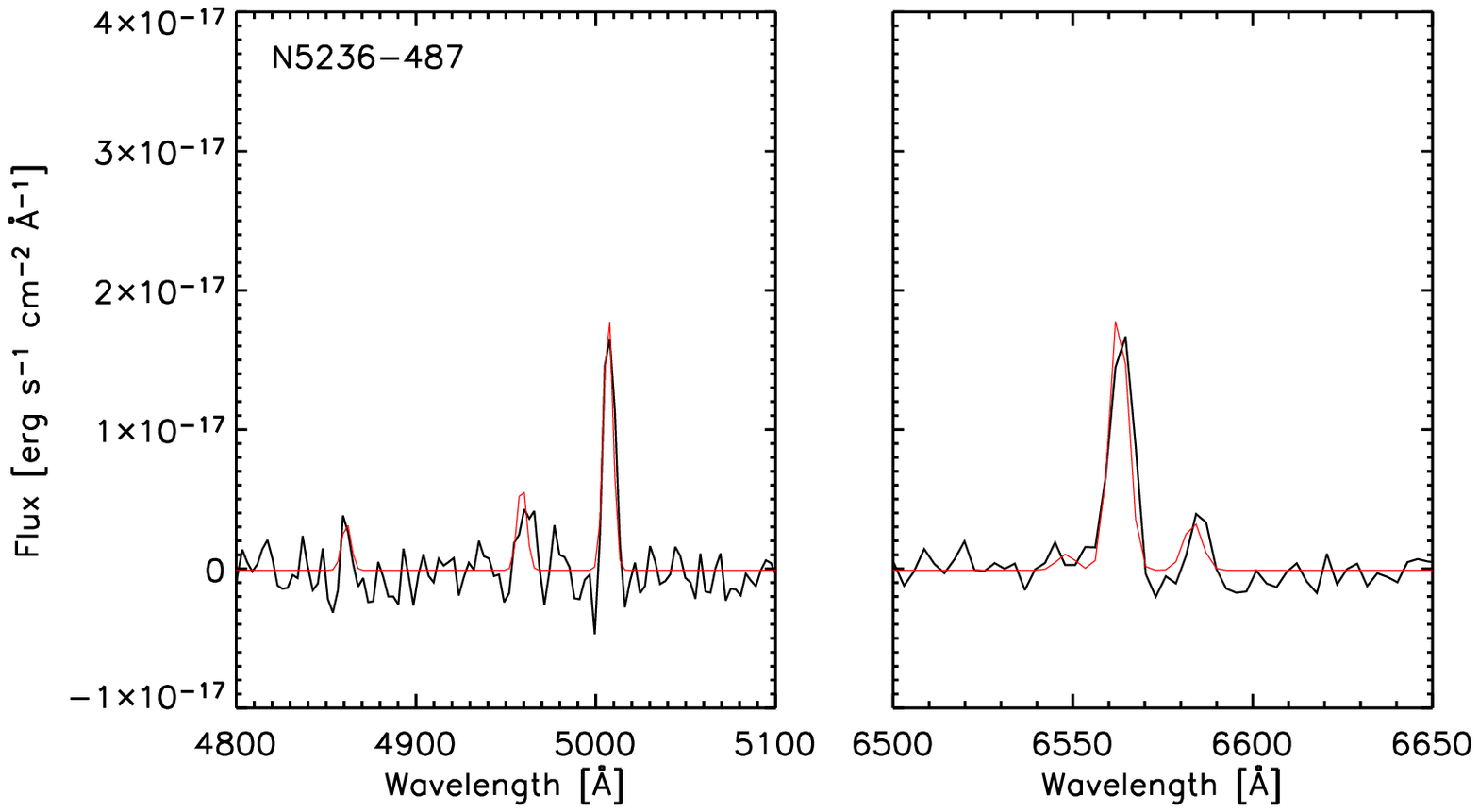}
\includegraphics[width=85mm]{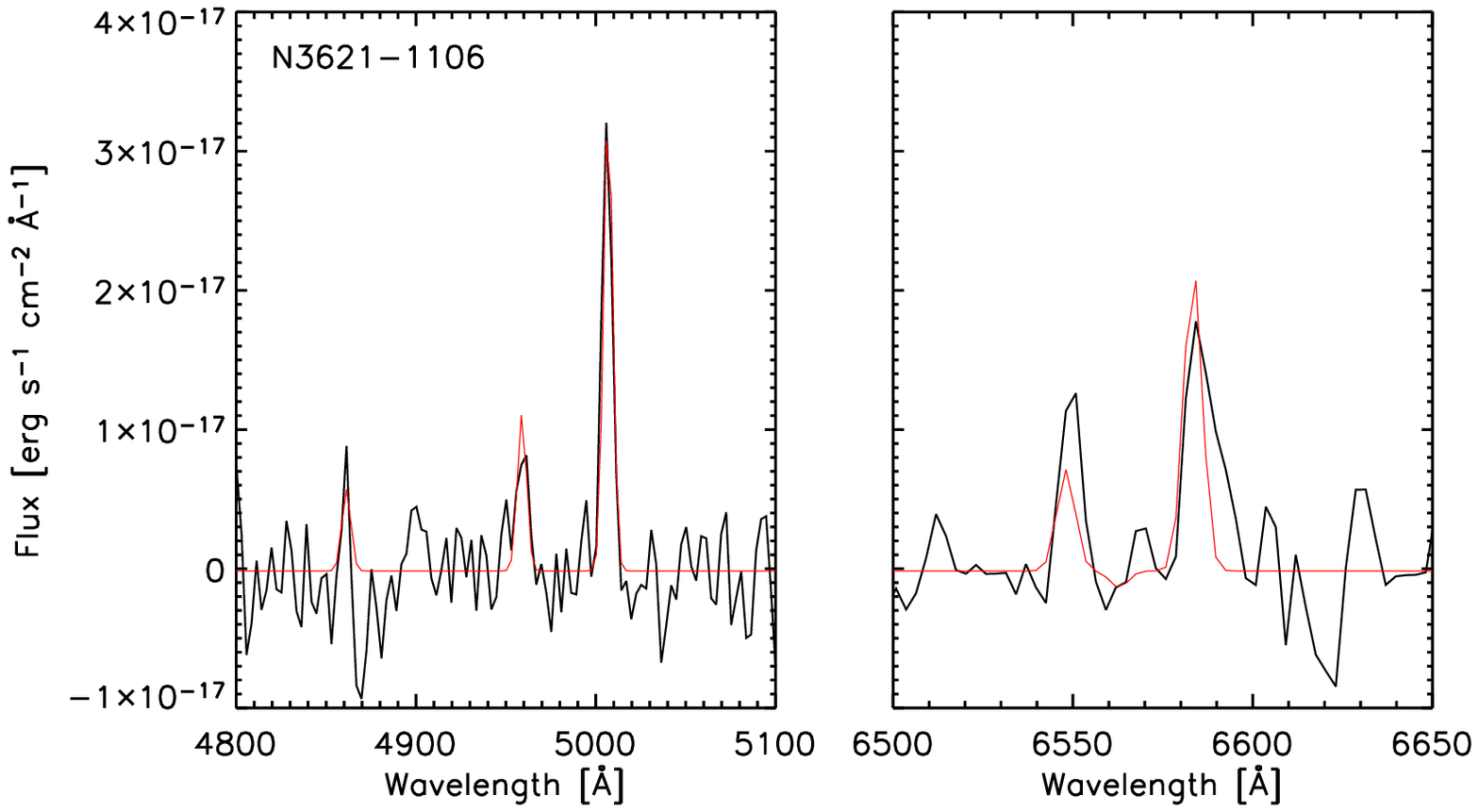}
\caption{Gaussian emission line fits to the spectra after subtraction of 
 best-fitting solar-metallicity SSP models. In each panel, the thick lines 
 represent the observed 
 spectra while the red lines are the 
 Gaussian fits. The regions
 around H$\beta$+[\ion{O}{iii}] and H$\alpha$+[\ion{N}{ii}] are shown
 on the same flux scale.}
\label{fig:emfit}
\end{figure}

Fig.~\ref{fig:emfit} shows the regions around the H$\beta+$[\ion{O}{iii}] and 
H$\alpha+$[\ion{N}{ii}] lines after subtraction of the best-fitting, scaled
solar-metallicity models. Emission line fluxes were measured by fitting
Gaussian line profiles to the emission line spectra. The
[\ion{O}{iii}] $\lambda$5007/4959\AA\ and [\ion{N}{ii}] $\lambda$6584/6548\AA\ 
line ratios were kept fixed at 3.0 as dictated by 
atomic physics (e.g.\ Osterbrock \cite{osterbrock}), while the 
[\ion{O}{iii}]/H$\beta$ and [\ion{N}{ii}]/H$\alpha$
ratios were allowed to vary. The resulting best-fitting Gaussian models
are overplotted on the spectra in Fig.~\ref{fig:emfit}. 
For the two emission-line sources associated with YMCs in NGC~5236, no 
further shifts in wavelength were required 
to match the line profiles and the sources are likely to be physically
associated with the clusters. The best fit for N3621-1106 was
obtained if an additional shift of about $+150$ km/s was applied to
the observed spectrum, so in this case the source of the emission lines
may not be physically associated with the cluster. 

Table~\ref{tab:emlines} lists the emission line fluxes obtained from 
the fits in Fig.~\ref{fig:emfit}. Errors were estimated by fitting
emission lines at 8 different wavelengths where none were expected.
The mean errors in the table represent the standard deviation of
these 8 test measurements. 
It should be emphasised that these error estimates only account for 
random measurement errors. The
line fluxes may be further affected by systematic errors, for example due to 
inaccurate subtraction of the stellar continuum. To investigate this issue,
we repeated the emission
line measurements after subtracting $Z=0.008$ SSP models based on the 
Padua isochrones, and $Z=0.008$ and $Z=0.020$ models based on the
Geneva isochrones. The resulting emission line fluxes generally agreed with 
those listed in Table~\ref{tab:emlines} within the stated one-sigma errors. 
Only for the H$\beta$ line in the spectrum of N5236-254 was a somewhat larger
deviation encountered -- for this line, a flux of 
$2.5\times10^{-17}$ erg$^{-1}$ cm$^{-2}$ was returned when using the
$Z=0.008$ Padua-based SSP model, i.e.\ a 2.7$\sigma$ deviation. In 
conclusion, although some systematic errors cannot be entirely ruled out,
the error estimates in Table~\ref{tab:emlines} appear to be fairly 
realistic overall. The flux calibration may account for
a further $\sim10-20$\% uncertainty on the line fluxes, although relative
line ratios are probably more accurate.
Finally, the $m_{5007}$ magnitudes, defined
as $m_{5007} = -2.5 \log(F_{5007}) - 13.74$ (Jacoby \cite{jac89}),
are also listed. 

\subsection{Host cluster properties} 
\label{sec:hostprops}

\begin{table}
\caption{\label{tab:phot}Cluster data. Photometry is corrected for foreground
  reddening only. $\Delta A_B$ is the additional amount of reddening
  required to best match the SSP model colours. The ranges of turn-off
  and total cluster masses reflect the differences between spectroscopic
  and photometric age estimates. The masses are for a Salpeter-like
  IMF with a lower mass limit of $0.1 M_\odot$}
\begin{tabular}{lccc} \hline
                         & N5236-254        &  N5236-487       &   N3621-1106 \\
$V_0$                    & $18.20\pm0.01$ & $19.96\pm0.04$ & $19.67\pm0.14$ \\
$(U-B)_0$                & $0.16\pm0.02$  & $-0.33\pm0.05$ & $-0.13\pm0.15$ \\
$(B-V)_0$                & $0.28\pm0.01$  & $ 0.20\pm0.05$ & $ 0.19\pm0.17$ \\
$(V-R)_0$                & $0.22\pm0.01$  & $ 0.27\pm0.05$ & $ 0.24\pm0.21$ \\ 
$(V-I)_0$                & $0.46\pm0.01$  & $ 0.60\pm0.05$ & $ 0.57\pm0.23$ \\ 
log(age/yr)$_{\rm phot}$ & $8.56\pm0.03$  & $7.81\pm0.24$  & $ 8.21\pm0.45$ \\
$\Delta A_B$ [mag]       & $0.00\pm0.03$  & $0.45\pm0.29$  & $ 0.30\pm0.55$ \\ 
log(age/yr)$_{\rm spec}$ &    8.45      &   7.50     &   7.70     \\ 
$M/10^3 M_\odot$         &  376--495    &   29--51   &  94--187   \\
$M_{\rm TO}/M_\odot$     &  3.2--3.6    &  6.6--9.0  &  4.5--7.3  \\ \hline
\end{tabular}
\end{table}

\begin{figure}
\includegraphics[width=85mm]{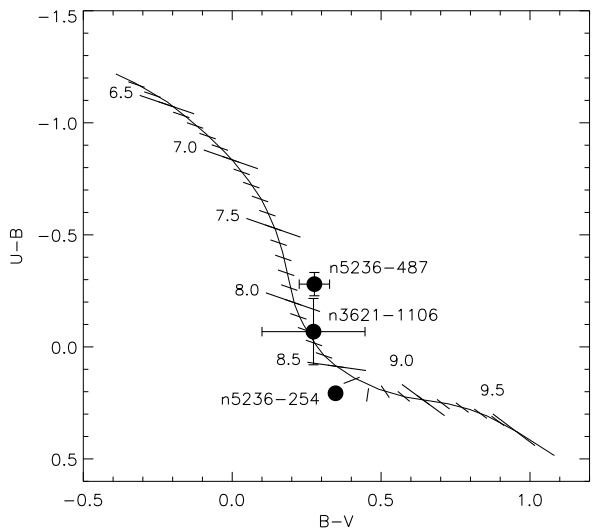}
\caption{\ub\ vs.\ \bv\ two-colour diagram for the three YMCs. Also shown
  is the S-sequence from Girardi et al.\ (1995). The numbers along the
  S-curve indicate the log(age) corresponding to the major tick marks.
  }
\label{fig:ubv}
\end{figure}

Photometry for the clusters was taken from Larsen (\cite{lar99}). 
Table~\ref{tab:phot} lists the $UBVRI$ broad-band colours for each
cluster, corrected for foreground extinction only. From the broad-band
colours, rough age estimates can be obtained by comparison 
with predictions by simple stellar population (SSP) models, or by using
the empirical `S'-sequence calibration (based on LMC clusters)
by Girardi et al.\ (\cite{gir95}). Fig.~\ref{fig:ubv} shows the
\ub\ vs.\ \bv\ two-colour diagram, including the S-sequence and photometry
for the three clusters. Based on this plot, N5236-487 appears to be
the youngest of the three objects with $\log$(age)$\sim$7.8, while
N5236-254 is the oldest ($\log$(age)$\sim8.6$). Including all the
colours available in Table~\ref{tab:phot} and comparing with solar metallicity 
SSP models from Bruzual \& Charlot (\cite{bc03}),
the photometric age and extinction estimates listed in
Table~\ref{tab:phot} were obtained. Here, $\Delta A_B$ denotes the
extinction required to best match the observed colours \emph{in addition}
to the foreground extinction. Errors on the age- and extinction estimates
were estimated by a Monte-Carlo procedure: First, random offsets 
$\delta_{\rm col}$ were added to the observed colours, and the age- and
extinction were then re-derived. The offsets $\delta_{\rm col}$ were
drawn from a Gaussian distribution with standard deviation equal to the
photometric errors on each colour.  The errors on the ages and extinctions
were then estimated as the standard deviation of the individual
values obtained from 100 such experiments.  The ages derived
from the comparison with SSP models agree quite well with those inferred 
from Fig.~\ref{fig:ubv}. 

The extinction estimates in Table~\ref{tab:phot} can be compared with
those derived from the H$\alpha$/H$\beta$ line ratios in
Table~\ref{tab:emlines}. Assuming the Case B recombination
ratio F(H$\alpha$)/F(H$\beta$) = 2.85 (e.g.\ Osterbrock \cite{osterbrock}),
we derive $\Delta A_B=0.02\pm0.95$ mag for N5236-254 and 
$\Delta A_B=2.51\pm1.53$ mag for N5236-487.  These values are consistent
with the photometric $\Delta A_B$ values, although the errors are larger.
In the case of N3621-1106, 
both H$\alpha$ and H$\beta$ are essentially non-detections, and it is
not possible to derive a meaningful extinction estimate.

Also listed in Table~\ref{tab:phot} are the spectroscopic age estimates,
which are simply the ages of the best-fitting SSP model spectra.  The 
spectroscopic ages tend to be lower than the photometric ones, although for 
any individual cluster the difference between the two age estimates is not 
much larger than the $1\sigma$ error.  However, the relative ranking of the 
clusters remains the same.  A detailed comparison of photometric and 
spectroscopic ages, using the full sample, will be carried out in a 
separate paper.  Because the continuum shape was not fitted, no constraints 
on the extinction are available from the spectral fits.

Once the ages and reddenings are known, the cluster masses can be estimated
by combining the SSP model predictions for mass-to-light ratios with
the observed magnitudes and adopting the distances for the galaxies
quoted above. Table~\ref{tab:phot} lists these photometric mass estimates,
assuming a stellar mass function with the Salpeter slope 
($dN/dM \propto M^{-2.35}$) extending down to 0.1 M$_{\odot}$. For a
more realistic IMF (e.g.\ Kroupa \cite{kroupa02}) the masses would be
lower by a factor of $\sim$1.6. The mass ranges given in the table
correspond to the difference between the photometric and spectroscopic
ages, with the lower masses corresponding to the younger ages.

The main sequence turn-off masses, $M_{\rm TO}$, corresponding to the cluster 
ages
were obtained from the Salasnich et al.\ (\cite{sal00}) stellar model
tracks. Again, Table~\ref{tab:phot} lists a range corresponding to
the photometric and spectroscopic ages. If the emission line sources
are indeed PNe associated
with their putative host clusters, then their progenitor stars should have
started out with masses close to the cluster turn-off masses.

\subsection{The nature of the sources: planetary nebulae?}

Several classes of objects show emission line spectra, but not
all are consistent with the characteristics of the objects studied here.
First, \ion{H}{ii} regions are common in star-forming galaxies.  Our full
sample contains several such objects, which generally show very different
spectra than those in Fig.~\ref{fig:spec}. The emission line equivalent 
widths of \ion{H}{ii} regions are generally much larger than in the spectra 
in Fig.~\ref{fig:spec}, and the excitation is typically
lower (e.g.\ Acker et al.\ \cite{acker87}). Furthermore,
young star-forming regions tend to have a much less regular structure than
the smooth profiles indicated by the images in Fig.~\ref{fig:stamps}.
Finally, only clusters younger than $\sim10$ Myrs contain stars which
are hot enough to produce the ionising radiation required to excite an
\ion{H}{ii} region, while we derive significantly higher ages for the
three objects studied here (\S\ref{sec:hostprops}). We can therefore exclude 
with fairly high
confidence that the objects are \ion{H}{ii} regions.

Wolf-Rayet stars also show strong emission lines, but again the spectra
are quantitatively (and qualitatively) very different from those observed 
here. In particular, the lines in W-R star spectra are very broad 
(several 1000 km/s) and would be easily resolvable at the spectral resolution 
of our EMMI data.  As for \ion{H}{ii} regions, the ages of the clusters
studied here make it highly unlikely that they contain W-R stars, whose
progenitors are believed to be massive ($>20 M_{\odot}$; Humphreys et al.\
\cite{hum85}) stars with
lifetimes of only a few Myrs.  We note, however, that a few objects in our full 
sample do show the characteristic W-R bumps at 4650\AA\ and 5812\AA\ 
(e.g.\ Torres \& Massey \cite{tm87}), but these features are entirely absent 
from the spectra in Fig.~\ref{fig:spec}.

A third possibility, which is more difficult to dismiss, is 
\emph{symbiotic stars} (SSs). These are binary 
systems in which a red giant is transferring material to a hot compact 
companion, such as a white dwarf
(Kenyon \cite{ken86}). Mass loss from the red giant results in a nebula
which is ionised by the hot companion, producing a spectrum very similar
to that of a planetary nebula. The main difference is that the
spectrum of a SS contains an extra component in addition
to the emission-line spectrum and the spectrum of the hot star, namely
that of the red giant. However, for our sources it is impossible to 
detect this small addition to the continuum on top of the cluster
spectrum.  On the basis of the data at hand, it is difficult to distinguish 
between symbiotic stars and planetary nebulae as the most likely sources
of the emission lines seen in our spectra. The progenitor stars are
also expected to be similar, in both cases involving red giants or AGB 
stars. 

Most surveys of Galactic and extra-galactic emission-like objects identify 
a larger number of PNe than SSs, and we might thus expect on statistical
grounds that our candidates are more likely to be PNe.
About 1500 PNe are known in the Milky Way (Kohoutek \cite{kohoutek01}), while 
less than 200 symbiotic stars are catalogued 
(Belcy{\'n}ski et al.\ \cite{bel00}). 
Unfortunately, the completeness of these catalogues is very difficult to
assess for either type of object, and the classification of at least 50
objects in the Milky Way alone is still ambiguous (Kohoutek \cite{kohoutek01}).
The total number of PNe in the Milky
Way is estimated to be about 25000 (Buzzoni et al.\ \cite{buz06}),
but this is little better than an order-of-magnitude estimate. The uncertainty 
on the number of SSs may be even greater, with estimates ranging 
from 3000 (Kenyon \cite{ken86}) to as many as 300.000 (Munari \& Renzini
\cite{mr92}).  

Surveys of PNe and SSs in the Magellanic Clouds have the
advantage that the distances are known with reasonable accuracy.
About a dozen SSs are known in the LMC and SMC. These have H$\beta$
luminosities in the range $7.7\times10^{33} - 5\times10^{34}$ erg s$^{-1}$
(Morgan \cite{morgan92}), i.e.\ all are fainter than the H$\beta$ 
fluxes derived for the objects studied here (\S\ref{sec:physprop}).
For typical effective temperatures of the hot component of $\sim10^5$ K 
(M{\"u}rset et al.\ \cite{muerset96}), the [\ion{O}{iii}] $\lambda$5007
flux is about an order of magnitude greater than the H$\beta$ flux. As
a crude estimate, there may then be about 10 SSs in the Clouds with 
L([\ion{O}{iii}]) $> 10^{35}$ erg s$^{-1}$ (although, again, an unknown
completeness correction may apply).  For comparison, more than 100 
PNe with L([\ion{O}{iii}]) $> 10^{35}$ erg s$^{-1}$ are catalogued in the 
Clouds (Jacoby et al.\ \cite{jac90}). Thus, at least at the high-luminosity
end, PNe appear to outnumber SSs by about an order of magnitude in
the Magellanic Clouds.  Furthermore, M{\"u}rset et al.\ (\cite{muerset96}) 
conclude that the LMC SSs most likely belong to a population older than 
$\sim$4 Gyrs, which is much older than the clusters studied here
(\S\ref{sec:hostprops}).  It is also worth noting that no SSs are 
known to be associated with stellar clusters in the Milky Way while about 
7 PNe have been found in Galactic globular or open clusters, as noted
in the Introduction.

We thus proceed under the assumption that the emission-line objects
found here are most likely planetary nebulae, although symbiotic stars
cannot be entirely ruled out as an alternative. Repeat observations
might help clarify this issue, as SSs often show significant variability
on time scales of days, weeks or months.  A detailed analysis of the
nebular properties using high-dispersion spectroscopy might also help
distinguish between SSs and PNe, since SSs typically have $\sim100$ times 
higher densities than PNe (Kenyon \cite{ken86}). As a further note, it
seems to us that the identification issue frequently receives less
attention than it merits in extragalactic PN surveys.

\subsection{Physical properties of the PN candidates}
\label{sec:physprop}

\begin{figure}
\includegraphics[width=85mm]{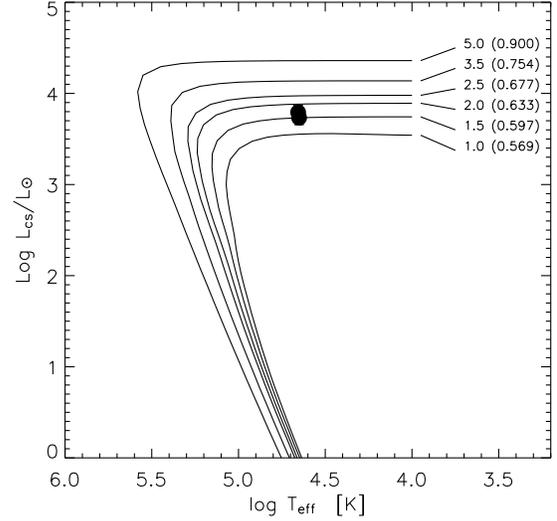}
\caption{H-R diagram for the PN candidates in NGC~5236, with 
  $\log$ T$_{\rm eff}$ and
  $\log L_{\rm cs}/L_{\odot}$ estimated from the $[$\ion{O}{iii}$]$ and
  H$\beta$ lines using the calibrations given by Dopita et al.\ (\cite{dop92}).
  Solar-metallicity post-AGB hydrogen-burning evolutionary tracks from 
  Vassiliadis \& Wood (\cite{vw94}) are shown for comparison, with the main 
  sequence and core mass (in parentheses) at $10^4$ K indicated. Both 
  objects are consistent with a core mass close to 0.60 M$_\odot$.
  }
\label{fig:hr_h}
\end{figure}

\begin{table*}
\caption{\label{tab:derived}Derived nebular properties. 
$L_{\rm cs}$ is the luminosity of the central star.  Note that H$\beta$
fluxes are derived from the H$\alpha$ fluxes assuming standard case B
recombination.}
\begin{tabular}{lccc} \hline
                              &  N5236-254    & N5236-487     & N3621-1106 \\
L (H$\alpha$) [$10^{35}$ erg s$^{-1}$]& 5.14$\pm$0.85 & 4.44$\pm$0.73 &  - \\
L (H$\beta$) [$10^{35}$ erg s$^{-1}$]$^\star$ & 1.80$\pm$0.30 & 1.56$\pm$0.25 &  - \\
L ([O III])                   & 6.11$\pm$0.99 & 5.10$\pm$0.84 & 18.7$\pm$4.49 \\
M$_{5007}$                   & $-3.01\pm0.18$ & $-2.82\pm0.18$ & $-4.23\pm0.26$ \\
Excitation parameter ($E$)    & 1.52$\pm$0.14 & 1.47$\pm$0.14 &  - \\
$\log T_{\rm eff}$ [K]        & 4.66$\pm$0.02 & 4.65$\pm$0.02 &  - \\
$\log L_{\rm cs}$ [$L_{\odot}$]   & 3.79$\pm$0.08 & 3.73$\pm$0.08 &  - \\
\hline
\end{tabular}
\end{table*}

Because only the strongest emission lines can be measured in the spectra,
it is not possible to derive basic nebular properties such as temperature
and density directly. While this limits the amount of physical information
that can be extracted from the spectra, some constraints can be put
on the total luminosities and effective temperatures of the ionising sources.
Dopita et al.\ (\cite{dop92}) have published an extensive set of
model calculations, allowing estimates of these properties from observations
of the Balmer lines and strong forbidden lines. They define the excitation 
parameter as $E = 0.045 \{ F(5007)/F(H\beta)\}$ for $0.0 < E < 5.0$.
For higher excitation parameters the \ion{He}{ii} $\lambda$4686\AA\ line
is required. This line is not detected in our spectra, but as will be
seen below the excitation parameters derived here are sufficiently low
that only the O and H lines are needed.

Measurements of the Balmer emission lines 
rely critically on correct subtraction of the underlying cluster spectra.
From Fig.~\ref{fig:emfit} it is difficult to assess the reliability
of this subtraction for H$\beta$. The H$\alpha$ line, on the other hand,
is stronger and less diluted by the stellar absorption, and therefore 
easier to measure reliably, at least for the two
objects in NGC~5236. 
For the object in NGC~3621, neither H$\beta$ nor
H$\alpha$ can be reliably measured. Thus, in the following discussion we
concentrate on the two objects in NGC~5236, for which the association
with the star clusters is anyway more secure.

Table~\ref{tab:derived} lists the basic parameters derived for the nebulae.
The H$\beta$ fluxes were estimated from the measured H$\alpha$ values
assuming the Case B recombination ratio,
but the results do not change much if we use
the measured H$\beta$ fluxes directly. For N5236-254, the de-reddened
H$\beta$ flux [$(1.80\pm0.47)\times10^{35}$ erg s$^{-1}$] agrees almost 
exactly with that derived from the H$\alpha$ flux.  For N5236-487 the 
de-reddened H$\beta$ flux is $(0.95\pm0.37)\times10^{35}$ erg s$^{-1}$, 
about 40\% lower than the value derived from H$\alpha$, but with a larger
formal error.
All line fluxes 
have been de-reddened using the Cardelli et al.\ (\cite{car89})
extinction curve and the extinction estimates from Table~\ref{tab:phot}.
The quoted error estimates include a 0.15 (0.20) mag
uncertainty on the distance modulus of NGC~5236 (NGC~3621).  For
reference, the [\ion{O}{iii}] luminosities have been converted to
absolute $M_{5007}$ magnitudes. These are well within the range 
expected for planetary nebulae; the tip of the planetary nebula luminosity
function is at $M_{5007} \approx -4.5$ (e.g.\ Kwok \cite{kwok00}).

Based on the excitation parameters and the H$\beta$ luminosities, the
effective temperatures and total luminosities of the central stars 
($L_{\rm cs}$) were 
derived using the relations given in Dopita et al.\ (\cite{dop92}).
The errors on $T_{\rm eff}$ and $L_{\rm cs}$ were estimated simply by
propagating the random measurement errors, and thus do not include any
systematic errors in the calibration. Such errros might result, for example,
from metallicity differences between the LMC and SMC calibrator PNe used
by Dopita et al.\ and the objects studied here. For the relevant
temperature range, Dopita et al.\ find that the 
L(H$\beta$)/L$_{\rm cs}$ ratio changes by only $\sim1$ percent for
O abundances between 0.1 and 2 times Solar (their Fig.~1). Over the same
metallicity range, $\log T_{\rm eff}$ varies by $\sim0.05$ dex for
a fixed $E\la2$. The error estimates in Table~\ref{tab:derived} do not
include uncertainties on the reddening.  For $E$ and $T_{\rm eff}$
the effect on the error budget is negligible ($\Delta \log T_{\rm eff} = 0.01$ 
mag for N5236-487) while the additional uncertainty on $L_{\rm cs}$ is
$\Delta \log L_{\rm cs} = 0.06$ dex for N5236-487. In both cases, the
associated errors for N5236-254 are even smaller.

The H-R diagram for the two nebulae in NGC~5236 is shown in 
Fig.~\ref{fig:hr_h} together with H-burning post-AGB evolutionary tracks
from Vassiliadis \& Wood (\cite{vw94}).  
Curiously, the two central stars
have very similar luminosities and effective temperatures, in both cases
corresponding to
a mass close to 0.60 $M_{\odot}$. This is somewhat
lower than predicted by the Vassiliadis \& Wood tracks for 
initial masses of 3.2--3.6 $M_{\odot}$ 
(for N5236-254)
and 
$>6.6 M_{\odot}$ (N5236-487). 
Empirically, the initial-final mass relation for
AGB and post-AGB evolution shows a fairly large scatter.  It may be nearly 
flat for $M_{\rm initial} < 3-4$ M$_\odot$,
although most studies suggest that a star with an initial mass 
$>6.6 M_{\odot}$ is more likely to produce a $\sim0.8-1.0$ M$_{\odot}$ 
remnant (Kwok \cite{kwok00}; Ferrario et al.\ \cite{fer05}; 
Weidemann \cite{we00}). It would
clearly be desirable to put stronger constraints on the physical
properties of these PNe.

We have used the solar-metallicity
(Z=0.016) tracks in Fig.~\ref{fig:hr_h}, but the Z=0.004 tracks do not
differ much and the inferred core masses would be lower by only $\sim0.02$
M$_{\odot}$.
The formal errors from Table~\ref{tab:derived} are smaller than 
the plot symbols, but we emphasise 
that a more robust analysis, using better data, would be desirable. 
However, since the 
model tracks are nearly horizontal in the H-R diagram, uncertainties on 
$T_{\rm eff}$ are of minor importance for the comparison with the post-AGB
tracks.
Likewise, uncertainties on the flux calibration and reddening
correction should lead to shifts of no more than 10-20\%, or $\sim0.1$
dex in $\log L_{\rm cs}$. The difference between the measured and
derived H$\beta$ flux for N5236-487 corresponds to a shift of
0.2 dex (downward) in $\log L_{\rm cs}$.

From Fig.~\ref{fig:emfit} one notes significant differences in 
the relative strengths of the [\ion{N}{ii}] and [\ion{O}{iii}] lines.
In N5236-254 and N3621-1106 the
[\ion{N}{ii}] lines are only slightly weaker than the [\ion{O}{iii}]
lines, while in N5236-487 they are about factor of 6 weaker.
This may appear somewhat puzzling, considering the very similar
parameters derived for the central stars in the two PNe in NGC~5236.
However, the PNe studied here fall well within the range of 
[\ion{O}{iii}]/[\ion{N}{ii}] line ratios seen in Galactic PNe 
(Cuisinier \cite{cui96}), which indeed
show a very large range in the relative abundances
of N and O. This is likely related to CNO processing in the progenitor stars 
(Peimbert \& Serrano \cite{peim80}) and the ratio of N to O is expected 
on theoretical grounds to increase steeply with 
progenitor mass (Renzini \& Voli \cite{ren81}). With the current data 
we cannot carry out an actual abundance analysis, but it may
be worth noting that the spectrum having the \emph{weakest} N lines is that
associated with the youngest cluster (and thus the \emph{highest} progenitor
star mass).
The H$\alpha$ line is also stronger relative to the [\ion{N}{ii}]
doublet in N5236-487 than in the other objects.  At least qualitatively, 
this seems consistent with the overall decrease in the [N/O] 
abundance ratio with decreasing [\ion{N}{ii}]/H$\alpha$ line ratio
noted by Perinotto \& Corradi (\cite{pc98}), again hinting that
N5236-487 may have a decreased [N/O] ratio, rather than an enhanced one
as expected for a massive progenitor.

\subsection{Observed vs.\ predicted number of PNe}

\begin{figure}
\includegraphics[width=85mm]{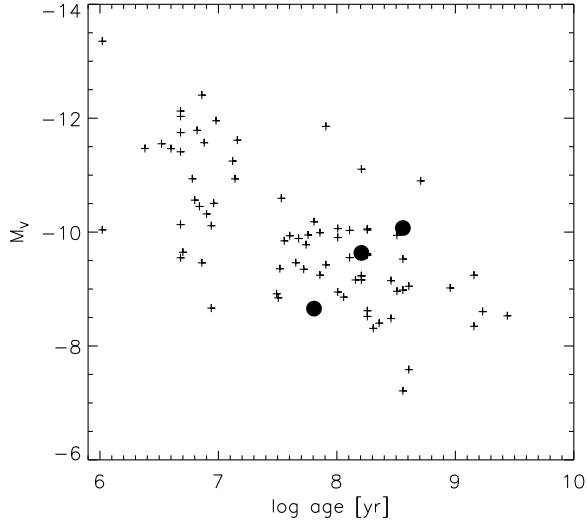}
\caption{Absolute $M_V$ magnitude versus log(age) for all clusters in
  the spectroscopic sample. The clusters containing PN candidates are
  indicated with filled circles. Ages in this figure are estimated
  from broad-band colours.}
\label{fig:age_mv}
\end{figure}

\begin{figure}
\includegraphics[width=85mm]{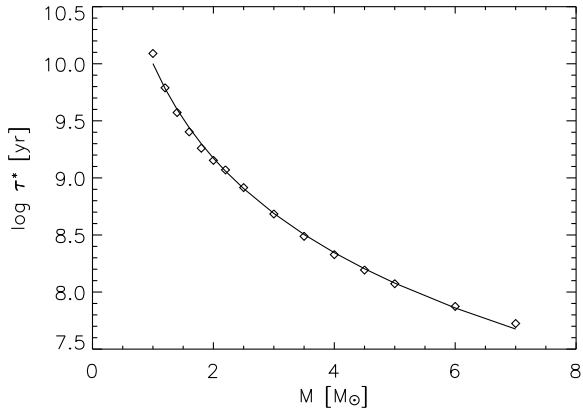}
\caption{Stellar lifetime as a function of initial mass, according to
  Salasnich et al.\ (\cite{sal00}) model tracks. The curve is a 
  power-law fit to the points.}
\label{fig:t_m_fit}
\end{figure}

\begin{figure}
\includegraphics[width=85mm]{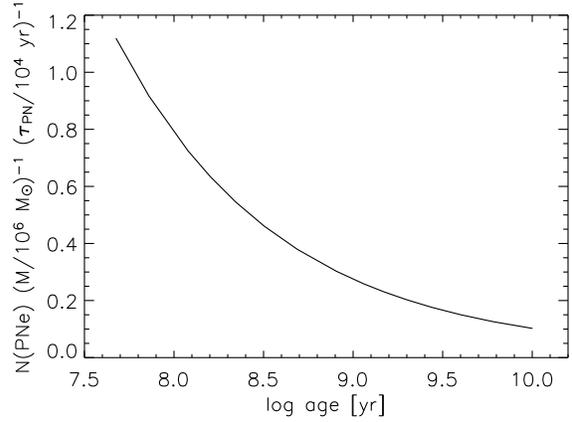}
\caption{Predicted number of planetary nebulae per $10^6$ M$_{\odot}$ 
  versus age for a single-aged stellar population, assuming a PN lifetime 
  of $10^4$ years, a Salpeter-like IMF, and the stellar mass-lifetime
  relation from Fig.~\ref{fig:t_m_fit}.}
\label{fig:npne}
\end{figure}

\begin{figure}
\includegraphics[width=85mm]{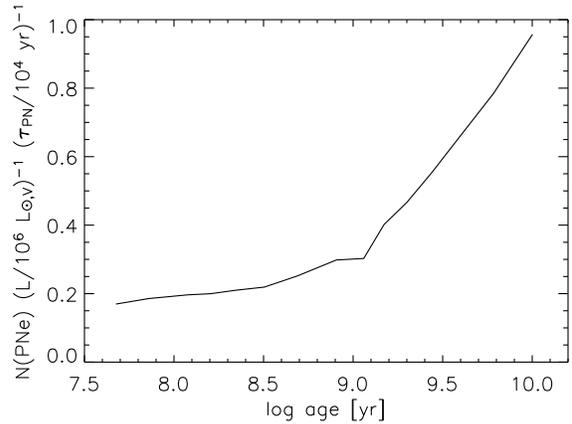}
\caption{Predicted number of planetary nebulae per $10^6$ L$_{\odot,V}$ 
  versus age for a single-aged stellar population, using the
  same assumptions as in Fig.~\ref{fig:npne} and mass-to-light ratios
  from Bruzual \& Charlot SSP models.}
\label{fig:npne_mv}
\end{figure}

Fig.~\ref{fig:age_mv} shows the absolute magnitude $M_V$ vs.\ the 
(photometric) age estimate for all 80 clusters included in the full 
spectroscopic sample. Ages were estimated by matching the observed $UBVRI$
colours against SSP model predictions from Bruzual \& Charlot
(\cite{bc03}) as described in \S\ref{sec:hostprops}.
Of the 80 clusters, 28 have ages less than 
$\log(t/yr)=7.5$ and are thus unlikely to host planetary nebulae. For the 
remaining 52 clusters, it is of interest to compare the expected number 
of PNe with the 2--3 objects actually detected (in the clusters marked 
by filled circles).

Recently, Buzzoni et al.\ (\cite{buz06}) have given an extensive discussion
of planetary nebulae as population tracers. However, their main focus
is on populations older than about 1 Gyr and their results are not
directly applicable to our case. 
In the following it is assumed that a stellar cluster consists of stars 
born in a single burst of negligible duration compared to the age of 
the cluster. As the age spread within a cluster is typically
only a few Myrs at most, this is a reasonable assumption for clusters 
which are old enough to host PNe. If the (initial) mass of a TP-AGB star
which is just about to become a PN is labelled $M_P$, then the stellar
death rate is the number of stars with masses in a small range 
($M_P..M_P+dM$), divided by the difference in the stellar lifetime 
($\tau^\star$) across the mass interval $dM$.  The predicted number 
of PNe in the cluster is then
\begin{equation}
  N_{\rm PN, pred.} = f_{\rm PN} \tau_{\rm PN} 
    \frac{N^\star(M_P + dM) - N^\star(M_P)}
         {\tau^\star(M_P) - \tau^\star(M_P+dM)},
  \label{eq:eq1}
\end{equation}
where $f_{\rm PN}$ is the fraction of stars which will actually end their 
lives as PNe and $\tau_{\rm PN}$ is the observable PN lifetime. $N^\star(M)$ is 
the number of stars in the cluster with masses less than $M$ and 
$\tau^\star(M)$ 
is the lifetime of a star with mass $M$. Eq.~(\ref{eq:eq1}) can be rewritten 
as
\begin{equation}
  N_{\rm PN, pred.} = 
   f_{\rm PN} \, \tau_{\rm PN} \, \left| \frac{M_{\rm cl} \, \xi (M_P)}{(d\tau^\star/dM)(M_P)} \right|
  \label{eq:npne}
\end{equation}
where $\xi$ is the stellar IMF, $\xi(M) = dN^\star/dM$, (normalised to a 
total mass of 1) and $M_{\rm cl}$ is the total mass of the cluster
(see also Renzini \& Buzzoni \cite{rb86} and Buzzoni et al.\ \cite{buz06}).  
For a Salpeter (\cite{salp55})
IMF populated with masses between $M_{\rm min}$ and $M_{\rm max}$ we have
\begin{equation}
  \xi(M) \, = \, 
    \frac{\alpha+2}{M_{\rm max}^{\alpha+2} - M_{\rm min}^{\alpha+2}}
    \,
    M^\alpha
\end{equation}
with $\alpha=-2.35$.  The denominator in Eq.~(\ref{eq:npne}) can
be estimated from stellar evolutionary tracks.
Fig.~\ref{fig:t_m_fit} shows the stellar lifetime until the TP-AGB phase as a 
function of initial mass
according to the evolutionary tracks in Salasnich et al.\ (\cite{sal00}).
Also shown in the figure is a power-law fit to $\tau^\star$ vs. $M$,
which gives $\tau^\star = 1.00\times10^{10} \, (M/M_{\odot})^{-2.75}$ yr.
Inserting the derivative of this fit in Eq.~(\ref{eq:npne}), the curve in 
Fig.~\ref{fig:npne} is obtained. This shows the predicted number of PNe
per $10^6$ M$_{\odot}$ as a function of age, assuming that
all stars go through the PN phase ($f_{\rm PN} = 1$) and that the PNe 
are observable for a period of $\tau_{\rm PN} = 10^4$ years 
(e.g.\ Marigo et al.\ \cite{mar01}), independent of initial mass.
The lifetime $\tau_{\rm PN}$ is probably the largest contributor to the
uncertainty in this calculation.  We also note that the power-law fit tends
to underestimate the slope at the low-mass end, so that the number of
PNe is over-predicted for ages greater than several Gyrs. However, for the 
age distribution of our clusters this is not a problem.

Using mass-to-light ratios from the Bruzual \& Charlot SSP models, the number 
of PNe per unit \emph{luminosity} can be calculated. This is shown in 
Fig.~\ref{fig:npne_mv} for the $V$-band.  While the expected frequency of 
PNe still depends on age, the dependence on assumptions about the
low-mass end of the IMF disappears when 
normalising to luminosity instead of mass. This is because the PNe progenitors 
are 
among the most luminous stars in a SSP while the low-mass stars contribute 
very little to the luminosity of a SSP. This makes the curve in 
Fig.~\ref{fig:npne_mv} more directly comparable to observations.
Of course, the same IMF has to be used when computing the number of 
PNe per mass (Eq.~\ref{eq:npne}) and for the subsequent evaluation of
the mass-to-light ratios. Here we are assuming a Salpeter-like IMF 
extending down to 0.1 $M_\odot$. The PN formation rate derived here
agrees within about a factor of two with the estimates by Renzini \&
Buzzoni (\cite{rb86}).

  We can now estimate the number of PNe expected in each cluster in 
Fig.~\ref{fig:age_mv}.
From Fig.~\ref{fig:npne} and Fig.~\ref{fig:npne_mv} it 
is clear that, on average, less than 1 PN is expected per cluster. By 
summing up the expected numbers of PNe in all clusters in the sample we 
estimate a total of 6 PNe in the age range $7.5 < \log({\rm age}) < 9.0$. This
is to be compared with the 2--3 objects actually observed. Note that
this assumes an observable PN lifetime of $10^4$ years -- if, for example, 
the lifetimes are shorter by a factor of two, then the predicted number
of PNe would be lower by the same factor. Since the PN lifetimes are a
steep function of the central star mass, with more massive progenitor
stars expected to produce shorter lifetimes (Sch{\"o}nberner \& Bl{\"o}cker 
\cite{sb96}), the agreement appears 
satisfactory and these rough estimates certainly suggest that the detection 
of a small
number of PNe in a spectroscopic survey like the one carried out here
is indeed expected.  Alternatively, we can compare with the empirical
estimate of the PN specific frequency from Magrini et al.\ (\cite{mag03}),
who estimate an average of one PN per $10^{6.92}$ $V$-band L$_\odot$ for
Local Group galaxies (ignoring any age dependence). The integrated $V$-band 
luminosity of all clusters in our sample older than 30 Myrs is about 
$3.3\times10^7$ L$_\odot$, corresponding to 4 PNe.  A search of PNe in a 
larger sample of young and intermediate-age star clusters might prove 
rewarding.

\section{Summary and conclusions}

In a sample of optical spectra of 80 young extragalactic star clusters,
3 emission line objects were detected which we identify as
likely PN candidates. Two of these have radial velocities
consistent with those of their host clusters, while the third displays
a shift of about 150 km/s and may be a field object falling on the
slit by chance. The analysis of the candidate PN spectra is 
complicated by the
relatively low spectral resolution and the strong underlying stellar spectra. 
This is a problem especially for the Balmer lines, which are very strong
in absorption in the cluster spectra and thus mask the comparatively
weak line emission from the PNe.  However, it was found that the line 
emission can be reasonably well separated by subtracting out synthetic SSP 
model spectra from the library of Gonz{\'a}lez-Delgado (\cite{gon05}).

All three objects have [\ion{O}{iii}] emission line fluxes consistent with 
those expected from
PNe. For the two objects where the association with clusters is most
robust (both 
in NGC~5236), the excitation parameters and total luminosities were estimated 
and the central stars were placed on the H-R diagram. Comparison with
post-AGB evolutionary tracks from Vassiliadis \& Wood (\cite{vw94}) 
indicates central star masses of about 0.60 $M_\odot$.

Based on stellar evolutionary timescales, we estimate that a total of 
about 6 PNe are expected in the 80 star clusters 
for a PN lifetime of $10^4$ years. Instead, 2--3 are detected
here, which we consider reasonable agreement. Scaling the
empirical estimate of the PN specific frequency from Magrini et al.\
(\cite{mag03}) predicts a total of 4 PNe in our sample, although
the age distribution of our PN host clusters is likely different from
that of the Local Group stellar populations on which the Magrini et al.\
estimate is based.  A systematic
search for PNe in a larger sample of extragalactic star clusters might
put useful constraints on the PN lifetimes and the range of stellar
masses that produce PNe. High-dispersion spectroscopy would allow
a better subtraction of the underlying cluster spectra and thus
more accurate measurements of the emission lines, and would
provide constraints on the kinematic properties (expansion velocities)
of the nebulae. Further observations would also provide
constraints on time variability, and thus help distinguish between
PNe and other possibilities (e.g. symbiotic stars).

\begin{acknowledgements}
We thank J.\ Walsh and A.\ Zijlstra for their encouragement and several
useful discussions. SSL thanks G.\ Pugliese for assistance and company 
during the observations. The anonymous referee provided several helpful
comments which led to significant improvements in the paper.
This research has made use of the NASA/IPAC Extragalactic Database (NED) 
which is operated by the Jet Propulsion Laboratory, California Institute of 
Technology, under contract with the National Aeronautics and Space 
Administration. 
\end{acknowledgements}

\end{document}